\documentclass{emulateapj}
\def\stacksymbols #1#2#3#4{\def\theguybelow{#2}
        \def\verticalposition{\lower#3pt}
        \def\spacingwithinsymbol{\baselineskip0pt\lineskip#4pt}
        \mathrel{\mathpalette\intermediary#1}}
\def\intermediary #1#2{\verticalposition\vbox{\spacingwithinsymbol
        \everycr={}\tabskip0pt
        \halign{$\mathsurround0pt#1\hfil##\hfil$\crcr#2\crcr
                \theguybelow\crcr}}}
\def\lta{\stacksymbols{<}{\sim}{2.5}{.2}}

\shorttitle{Mid-IR Surface brightness and colors in Elliptical Galaxies}
\shortauthors{Temi, Brighenti \& Mathews}

\begin{document}

\title{THE MID-INFRARED SPECTRAL ENERGY DISTRIBUTION, 
SURFACE BRIGHTNESS AND COLOR PROFILES IN ELLIPTICAL GALAXIES}

\author{Pasquale Temi\altaffilmark{1,2},
Fabrizio Brighenti\altaffilmark{3,4}, William
G. Mathews\altaffilmark{3} }
%\email{ptemi@mail.arc.nasa.gov}
%\email{mathews@ucolick.org}
%\email{fabrizio.brighenti@unibo.it}

\altaffiltext{1}{Astrophysics Branch, NASA/Ames Research Center, MS
  245-6,
Moffett Field, CA 94035.}
\altaffiltext{2}{SETI Institute, Mountain View, CA 94043;
and Department of Physics and Astronomy, University of Western
Ontario,
London, ON N6A 3K7, Canada. ptemi@mail.arc.nasa.gov}
\altaffiltext{3}{University of California Observatories/Lick
  Observatory,
Board of Studies in Astronomy and Astrophysics,
University of California, Santa Cruz, CA 95064 
mathews@ucolick.org}
\altaffiltext{4}{Dipartimento di Astronomia,
Universit\`a di Bologna, via Ranzani 1, Bologna 40127, Italy 
fabrizio.brighenti@unibo.it}

\begin{abstract}
We combine 2MASS data and {\it Spitzer} archival data
to study the emission in mid-infrared passbands (1.2 - 24$\mu$m)
from a sample of 18 elliptical galaxies.
In general the surface brightness distributions resemble
de Vaucouleurs profiles, indicating that most of the emission
arises from the photospheres or circumstellar regions of
red giant stars.
The spectral energy distribution peaks near the $H$ band
at 1.6$\mu$m.
The half-light or effective radius
has a pronounced minimum near the $K$ band (2.15$\mu$m) 
with a second, less consistent minimum in the 24$\mu$m passband.
All sample-averaged radial color profiles
$\langle \lambda_i - \lambda_j \rangle$ 
where $\lambda_i < \lambda_j$
(and $j \ne 24\mu$m) have positive slopes within about 
twice the ($K$ band) effective radius. 
Evidently this variation arises because of 
an increase in stellar metallicity toward the galactic cores.
Color profiles $\langle K - j \rangle$ all have positive slopes,
particularly when $j = 5.8\mu$m although no obvious absorption feature
is observed in spectra of elliptical galaxies near 5.8$\mu$m.
This, and the minimum in $R_e$, suggests that the $K$ band may be
anomalously luminous in metal-rich stars in galaxy cores.
Unusual radial color profiles involving the 24$\mu$m passband
may suggest that some 24$\mu$m emission comes from
interstellar not circumstellar dust grains.
\end{abstract}

\keywords{galaxies: elliptical and lenticular; galaxies: ISM;
infrared: galaxies; infrared: ISM}

\section{Introduction}

Our curiosity about the mid-IR properties of elliptical galaxies 
arose from two motivations: the implications for stellar mass 
and light distributions and the transition from circumstellar 
dust to interstellar dust.  
Regarding the structure of elliptical galaxies, 
we were surprised that the effective (or half-light)  
radius $R_e$, a fundamental structural property of elliptical galaxies,  
was a surprisingly strong function of wavelength. 
Between the $V$ and $K$ bands the effective radius $R_e$ decreases 
by more than 30 percent (Ko \& Im 2005), reducing the ``tilt'' of the 
fundamental plane for elliptical galaxies from that 
expected for virialized 
homologous galaxies (e.g. Prugniel \& Simien 1996).
Radial variations in the  
mean stellar color can arise from variations in either 
stellar metallicity or age. 
However, it is generally accepted that the observed 
color variations in elliptical galaxies 
are due largely to metallicity 
(e.g. Sanchez-Blaquez et al. 2007). 
Photometric colors are affected 
by an increase in the spectral density and strengths 
of absorption lines (line blanketing) 
toward the galactic centers where the stellar metal abundance 
reaches its maximum value. 
When imaged in the light of optical photometric bands,  
line blanketing reduces the stellar emissivity and surface brightness 
in the galactic cores, resulting in an overall increase 
in the half-light radius when compared to the 
surface brightness profiles in relatively line-free near-IR 
photometric bands such as the $K$ filter near 2$\mu$m.
If the $K$ band is essentially free of strong absorption and 
emission features, as commonly supposed, we wondered if 
the effective radius would continue to decrease at wavelengths 
greater than the $K$ band. 
More importantly, does the $K$ band emission, or the emission 
from any other infrared band, accurately trace the stellar 
luminosity and  
mass in elliptical galaxies, or is it necessary to make bolometric 
corrections using the full spectral energy distribution 
at each radius?

Our second motivation to study the mid-IR color profiles 
in elliptical galaxies was to learn more about the transition of 
infrared-emitting dust from circumstellar to 
interstellar environments. 
In a recent paper we showed that the total luminosity 
of elliptical galaxies in 
{\it Spitzer}'s 24$\mu$m photometric band correlated with optical 
luminosity while luminosities at 60 and 170$\mu$m did not 
(Temi, Brighenti, \& Mathews 2007). 
As dusty gas flows away from mass-losing red giant stars, 
it receives progressively less illumination from the  
photosphere of the parent star and relatively 
more collective diffuse radiation from distant galactic stars. 
Truly interstellar dust is heated at comparable rates by 
absorption of diffuse starlight and by inelastic impacts of thermal 
electrons in the hot interstellar gas. 
Both types of heating increase toward the galactic cores 
where the stellar and interstellar gas densities are largest. 
Does the 24$\mu$m emission from relatively cold 
dust in ellipticals vary with galactic 
radius in a manner that reveals 
a transition to interstellar heating patterns?  

We describe below the surface photometry 
of a sample of elliptical galaxies 
in the {\it Spitzer} photometric passbands at 
3.6, 4.5, 5.8, 8.0 and 24$\mu$m and compare them with 
similar 2MASS data at 1.2, 1.6 and the $K$ band at 2.15$\mu$m.
In general we find significant radial variations in colors formed by 
pairs of infrared bandpasses, particularly when the 
$K$ band is involved. 
The atomic, molecular or dust spectral features responsible 
for these
variations are currently unknown, but we discuss several 
possibilities. 

Typical spectra of elliptical galaxies taken with the 
{\it Spitzer} Infrared Spectrograph (IRS) 
(Bregman, Temi \& Bregman 2006; Bressan et al. 2007) 
have slope changes at about 8$\mu$m which may mark 
the transition from photospheric emission to emission from 
circumstellar dust, probably silicates (Bressan et al. 2007), 
but this transition is unlikely to be abrupt.  
In any case, photospheric gas emission and 
circumstellar dust emission must be regarded
as equally viable measures of stellar light in elliptical galaxies. 
It has been known for many years that 
the surface brightness profiles in these galaxies 
at 10-12$\mu$m follows a de Vaucouleurs law similar to that of 
the optical light (Knapp, Gunn \& Wynn-Williams 1992).

\section{The Sample}

We selected 18 elliptical galaxies from the {\it Spitzer}
archives, emphasizing those with E classifications rather than 
E/S0 or S0. 
An attempt was made to select normal galaxies free of unusual 
masses of acreted cold gas, recent mergers or other such 
anomalies.
We restricted the absolute magnitude to 
$M_B > -18$ to avoid dwarf ellipticals. 
Most sample galaxies had been photometrically observed 
in the $K$ band by Pahre (1999) and all but six 
were photometrically analyzed 
in the near-IR 2MASS survey (Jarrett 2003). 
As described below 
we photometrically extracted the $K$ band 
surface brightness distributions  
and other photometric data directly from the 2MASS images for 
all galaxies in our sample including by necessity the 
six with unpublished 2MASS data. 
All sample galaxies were chosen so that the $K$ band effective 
radius exceeds 13$^{\prime\prime}$, allowing observations 
%with the IRAC bands and MIPS 8 and 24$\mu$m bands 
to extend well beyond the point response function at 
all infrared wavelengths. 
%In summary, our sample listed in Table 1 is adequate for 
%our purposes here, but would not have been the one we would 
%have chosen {\it ab initio}.
Our sample galaxies are listed in Table 1. 

\section{Observations}

The data presented here were obtained with the Infrared Array
Camera (IRAC) (Fazio et al. 2004)
and the Multiband Imager Photometer (MIPS) (Rieke et al. 2004) on
board the {\it Spitzer} Space Telescope (Werner et al. 2004).
At all wavelengths full coverage 
imaging was obtained for all observations with additional sky
coverage to properly evaluate the background emission. 
We used the  Basic Calibrated Data (BCD) products from the Spitzer
Science pipeline (version 14.0) to construct mosaic
images for all objects. Pipeline reduction and post-BCD processing
using the MOPEX software package (Makovoz et al. 2006)
provide all necessary steps 
to process individual frames: dark subtraction,
flat-fielding, mux-bleed correction, flux calibration, correction
of focal plane geometrical distortion, and cosmic ray rejection.

Foreground stars and background galaxies were present in the final
mosaiced images at all bands. These were identified by eye
and cross-checked using surveys at other
wavelengths (Digital Sky Survey and 2MASS).
They were then masked out from each IRAC and MIPS image before
isophotal fitting procedure was performed.
Sky subtraction was performed by averaging values from multiple
apertures placed around the target, avoiding any overlap
with the faint extended emission from the galaxy.

When data are presented in the IRAC and MIPS magnitude
system relative to Vega, we used the zero-magnitude flux
densities presented in the instrument's manuals and
published by Reach et al. (2005). 
Corrections for extended emission were also applied
to the flux densities as described in the Spitzer Observer's 
Manual.
%Photometric errors listed in Table 2 (see below) 
%refer only to the
%statistical uncertainties. Total errors should include
%systematic uncertainties due to flux calibration and to
%sky subtraction errors.
The uncertainties on the final absolute calibration are
estimated at few percent at all 4 IRAC bandpasses and 10\%
for the 24$\mu$m data.
                          
\section{Surface Brightness Profiles}

Following a procedure similar to that 
used in the 2MASS survey (Jarrett 2003), 
our surface photometry is based on fitting elliptical apertures 
to galactic images in every infrared passband.
This was accomplished using the ELLIPSE task in the 
ISOPHOTE package of IRAF. 
Bad pixels, superimposed stars and background objects were removed.
We continue to the sky noise limit at which the mean brightness 
on the best-fitting ellipse is 
19-20 mag arcsec$^{-2}$ in all passbands 
except at 24$\mu$m where the sky 
cutoff is at 17-18 mag arcsec$^{-2}$.

Figure 1 (top panel) shows the surface photometry for NGC 4472, a typical 
bright elliptical in our sample, 
plotted against $R(^{\prime\prime})^{1/4}$ 
where $R$ is the radius of the local semimajor axis.
For comparison we also show the surface brightness profile 
$\mu_K(R)$ of an image in the 2MASS K$_s$ filter. 
The bandpass of the 2MASS K$_s$ filter (2.0 - 2.32$\mu$m) 
differs slightly from the traditional 
Johnson K filter (2.03 - 2.42$\mu$m; 
see Pahre 1999 for further details). 
For simplicity  
in the subsequent discussion we refer to the K$_s$ filter simply as K. 
The surface brightness profiles $\mu_i(R)$ 
at the IRAC photometric bands 
$i = 3.6$, 4.5, 5.8 and 8$\mu$m shown in Figure 1 
overlap but are bounded below by 
$\mu_K$ (solid line) and above by $\mu_{24}$ (dotted line).
While a de Vaucouleurs $R^{-1/4}$ model is a good fit for all 
mid-IR wavelengths, 
small radial color variations between $\mu_i(R)$ 
do occur and these are discussed below. 
Our full set of surface brightness data is available in electronic 
form at the ApJ website. 

The lower two panels in Figure 1 show the spectral energy
distributions (SED) for NGC 4472 in magnitude arcsec$^{-2}$ and in
mJy arcsec$^{-2}$ plotted between the J (1.2$\mu$m) and 
24$\mu$m MIPS bands. 
Small radial variations in the infrared SED are visible. 
The mean E galaxy infrared SED for all galaxies in our sample 
(arbitrarily normalized to the $K$ band) is shown in Figure 2.
Evidently the energy output of elliptical galaxies
peaks near the $H$ band (1.6$\mu$m), 
in good agreement with the theoretical expectations 
(e.g. Piovan et al. 2003). 
Note that the sample-averaged $\mu_{24}$ is enhanced toward the
galactic center relative to other infrared passbands (also see Figs. 6
and 7 below). Although this 24$\mu$m enhancement is less pronounced in
NGC 4472 (Fig. 1), it is a general property of our sample galaxies.

Two galaxies in our sample -- NGC 4697 and NGC 5322 -- 
have atypical SEDs in the central $\sim5$ arcseconds. 
The first of these ellipticals has an unusually massive central 
dusty disk which, according to Bregman, Bregman, and Temi (2006) 
contains strong PAH emission possibly associated with a starburst 
that may have occurred some time ago. 
A small central dust lane in NGC 5322 has been described 
by Carollo et al. (1997). 
Infrared emission from these two central dust clouds is 
confined to the nucleus and does not contribute significantly 
to the color profiles discussed below which begin at 
$\sim6$ arcseconds. 

\section{Effective Radius}

%\subsection{Photometric Procedure}

The total magnitude of each galaxy at each 
infrared wavelength $i$ can be 
found by integrating the surface brightness $\mu_i$ over area.
This has been done in a sky-corrected manner by 
extending the observed $\mu_i(R)$ with  
extrapolations to 6$R_e(K)$ assuming de Vaucouleurs profiles 
and uniform ellipticity in the extrapolated extensions. 
%The contribution of the extrapolation to the total galactic 
%flux is typically less than 15 percent.
The effective or half-light radius $R_e(i)$ at each 
infrared wavelength was then determined 
by integrating over the surface brightness to that radius 
$R_e(i)$ enclosing half the total flux (data plus extrapolation). 
Since our photometric procedure is similar to that used 
in the 2MASS survey (Jarrett 2003), our values of 
$R_e(K)$ determined from the 2MASS image should agree with 
2MASS $R_e(K)$ as tabulated by Jarrett 
and Figure 3 shows that this is indeed the case.

%\subsection{Results} 

Figure 4 shows the ratio $R_e(i)/R_e(K)$ for our sample galaxies 
plotted against $R_e(K)$. 
While there is some variation among the galaxies, 
$R_e(i)$ is systematically larger than $R_e(K)$ for all 
IRAC wavelengths 
except $i = 24$ where the scatter is larger. 
$R_e(J)$ and $R_e(H)$ also tend to exceed $R_e(K)$, 
indicating a minimum half-light radius in the $K$ band. 

Figure 5 shows sample-averaged values of $\langle R_e(i)/R_e(K)\rangle$ 
from $i = V$ to 24 microns. 
The leftmost open diamond showing $\langle R_e(V)/R_e(K)\rangle$ 
is based on $R_e(V)$ for each sample galaxy 
taken from Faber et al. (1989), 
corrected slightly for seeing by Pahre (1999) 
using a program developed by Saglia et al. (1993).
Circular aperture photometry is
generally used in the Faber et al. compilation 
and a fully consistent ellipsoidal determination of 
$R_e(V)$ would be slightly different.
For comparison the + symbol shows 
$\langle R_e(V)/R_e(K)\rangle = 1.33 \pm 0.40$ 
which is the mean ratio for all 273 early type galaxies in 
the Pahre (1999) sample (Ko \& Im 2005). 
Finally, in computing $\langle R_e(V)/R_e(K)\rangle$ 
for Figure 5 we excluded NGC 4696, the central 
and most luminous galaxy in the Centaurus Cluster. 
This galaxy has very extended stellar halo in the V band 
that must begin at a radius that exceeds the sky cutoff
for all the IR bands. 
The infrared properties of NGC 4696 appear normal in other 
respects although this galaxy has an optically visible 
dust land and is often classified as an S0.

Figure 5 clearly shows that the minimum effective 
radius occurs at the $K$ band, 
although $\langle R_e(24)/R_e(K)\rangle \approx 1$. 
The steady decrease in the effective radius 
from the optical to the $K$ band 
can be largely (or completely) 
explained by the decreasing influence of line blanketing 
toward the red.
The unanticipated increase in $R_e$ at 3.6 microns and beyond 
is likely to result from molecular and dust 
absorption as we discuss below. 

Table 2 lists values of the effective radius $R_e(i)$, 
the surface brightness at the effective radius 
$\mu_e(i) = \mu_i[R_e(i)]$, and bandpass luminosities $L_i$ 
for individual galaxies. 

\section{Radial Color Gradients}

Figures 6 and 7 show infrared color gradients 
$\mu(i) - \mu(j)$ averaged over our sample galaxies. 
This is done in several steps.
First the $\mu(i) - \mu(j)$ profile for each galaxy is plotted 
against physical radius 
normalized to the effective radius $R_e(K)$ for that galaxy. 
Then plots of $\mu(i) - \mu(j)$ for all sample galaxies 
are binned in 
$\Delta[R/R_e(K)]$ and averaged as shown in Figures 6 and 7.
Consequently, the error bars in Figures 6 and 7 
show the typical 1$\sigma$ dispersion of sample 
galaxies from the mean color profiles in each bin; 
they do not imply that the shapes of the profiles 
are uncertain by this amount since the profiles tend to be 
remarkably similar for most sample galaxies. 
Color profiles involving 24$\mu$m were scaled 
downward for comparison with other profiles. 
We show in the Appendix that the point spread functions 
in the $K$, IRAC, and 24$\mu$m passbands are small and do not 
substantially alter the surface brightness color variations 
in Figures 6 and 7. 

In general we see that infrared colors are quite sensitive to 
radial stellar population variations in elliptical galaxies. 
In principle, 
color gradients can arise from radial variations of stellar 
metallicity or mean 
stellar age, but recent detailed radial-dependent 
population models show that metallicity is by far the 
dominant source of color gradients (Sanchez-Blaquez et al. 2007).

Assuming that metallicity is the primary driver for the 
radial color variations in Figures 6 and 7, 
we attempted to compare mean colors in our sample galaxies with 
global metal abundances determined from optical line indices. 
For example we plotted all the colors discussed above averaged 
at $R_e(K)$, $R_e(K)/2$ and $R_e(K)/4$ against the 
metal abundances recently determined by 
Sanchez-Blazquez et al. (2006) who studied many of the 
galaxies in our sample. 
This resulted in scatter diagrams because the errors 
in the metal abundances of individual galaxies 
using optical line indices 
are comparable to the complete range of metal abundances 
in our sample.
Since the stellar metal abundance is known to 
increase with total luminosity, 
we also sought correlations of color-related parameters 
with the luminosity of 
each sample galaxy but, as before, no correlation emerged.

\section{Discussion}

Relatively little is understood about the detailed 
mid-IR spectra of metal (and oxygen) rich giant stars 
expected in elliptical galaxies; 
the most relevant observations may be the 
5 - 17$\mu$m spectra of Milky Way Bulge AGB stars 
(Blommaert et al. 2006).
However, it is well known from studies of 
globular clusters that the red giant branch cools with increasing 
metallicity (e.g. Ferraro et al. 2000), 
allowing molecular formation.  
Consequently, it is likely that molecular 
or dust spectral features contribute 
to the mid-IR spectrum and to the color variations 
we observe. 

The $\langle K - 24 \rangle$ 
and $\langle 4.5 - 24 \rangle$ color profiles 
in Figures 6 and 7 are quite peculiar 
so we begin by discussing the colors between other passbands. 
In particular we see in Figure 7 that 
colors formed among the 3.6, 4.5 and 8.0$\mu$m bands have the 
smallest slopes.
One might imagine therefore 
that these three bands are uniquely free of metal 
lines or other metal-dependent spectral features, 
but such a conclusion would probably be premature. 
Nevertheless it is remarkable that 
the colors in Figure 6 formed by combinations of these 
three passbands with the $K$ band -- $\langle K - 3.6 \rangle$,
$\langle K - 4.5 \rangle$, and $\langle K - 8.0 \rangle$ -- 
have much larger positive slopes 
than $\langle 4.5 - 8 \rangle$ or $\langle 3.6 - 4.5 \rangle$, 
so the $K$ band is sensitive 
to gradients in the stellar metal abundance. 
The sense of the positive slopes in Figures 6 for 
colors 
$\langle K - i \rangle$ where $\lambda_K < \lambda_i$ is either that 
the $i$ band is more strongly absorbed in the high metallicity 
core or the $K$ band has an emission feature there. 
Emission in the $K = K_s$ band (2.0 - 2.32$\mu$m) seems unlikely. 
The Brackett-$\gamma$ line (2.16$\mu$m) 
is not expected to appear strongly in emission 
when the Balmer lines themselves are weak or absent 
and the CO band heads 
longward of 2.29$\mu$m, if present in emission, affect only a small 
part of the $K_s$ band and are more likely to appear in 
absorption (e.g. Cunha \& Smith 2006). 
The colors in Figure 6 with the greatest positive slopes, 
$\langle K - 5.8 \rangle$ and $\langle K - 8.0 \rangle$, 
are interesting because the 
5.8$\mu$m ($4.97 - 6.47\mu$m) and 8.0$\mu$m (6.37 - 9.53$\mu$m) 
bandpasses are visible in 5-20$\mu$m IRS spectra taken with 
{\it Spitzer} (Bregman, Temi \& Bregman 2006; Bressan et al. 2007).
These spectra, which are in good agreement with IRAC photometry, 
show no visible systematic absorption near 5.8 or 8.0$\mu$m 
that would be needed to account for the positive slopes 
of $\langle K - 5.8 \rangle$ and $\langle K - 8.0 \rangle$
in Figure 6. 
However, it is possible that some broad band dust absorption 
is present at all IRAC wavelengths in the circumstellar regions of 
metal rich stars. 
If so, some enhanced solid phase absorption 
near 5.8$\mu$m would be needed to account for the larger 
slope of $\langle K - 5.8 \rangle$. 

In view of these unsatisfactory and somewhat contradictory 
arguments, we are led to speculate that the $K$ band 
may be largely free of intrinsic absorption (or emission) features 
in the metal rich cores of these galaxies but that 
considerable widely distributed line or dust absorption 
exists to the blue (and possibly also to the red) of the $K$ band. 
Under these circumstances, radiation at wavelengths 
adjacent to the $K$ band 
could be absorbed and reprocessed in the stellar atmospheres 
until it preferentially emerges in the $K$ band, 
resulting in an anomalously enhanced emission there.
This hypothesis is also consistent with Figure 5 in which the $K$ band 
has the smallest effective radius, i.e. its radiation is most 
strongly peaked in the central, most metal-rich regions 
of the galactic cores. 
An alternative, less likely, possibility for an enhanced 
$K$ band is emission from 
hot, newly formed dust grains having a restricted range of 
temperatures and emissivities that peak in the $K$ band. 

The strange non-monotonic color profiles for  
$\langle K - 24 \rangle$
and $\langle 4.5 - 24 \rangle$ in Figures 6 and 7 
may be related to an additional interstellar component 
in the 24$\mu$m emission. 
To explore the possibility that these cold grains are 
heated by diffuse starlight, we compared $(K - 24)$ with the 
local mean intensity of galactic stars.
Specifically, we evaluated $\langle K - 24 \rangle$ near 
$R = R_e(K)/2$ for each sample galaxy and 
compared this with $J_B[R_e(K)/2]$, 
the mean diffuse $B$ band intensity at this same radius 
found by using (spherical) de Vaucouleurs profiles and distances 
from Table 1 
(Temi, Brighenti \& Mathews 2007).
This comparison, shown in Figure 8, 
appears indecisive due to the large scatter. 
The two relatively low luminosity 
galaxies at the far right, NGC 584 and 3379, 
have larger $J_B$ because of the stronger stellar concentration 
in low luminosity ellipticals. 
While NGC 584 is classified as an E4 in RC2, 
Sandage \& Bedke (1994) argue that it is an S0. 
Many authors have claimed that NGC 3379 is  
a face-on S0 (Capaccioli et al. 1991; 
Statler \& Smecker-Hane 1999; see 
Harris et al. 2007 for a  recent discussion). 
NGC 584 is certainly a flat galaxy and if NGC 3379 is also 
a flat S0 galaxy  
viewed face-on, we may have overestimated $J_B$ 
of both galaxies by assuming that they are spherically symmetric. 
If these two galaxies are removed (or shifted to the left), 
the positive correlation visible in Figure 8 would be 
strengthened and this would be consistent with 
(stochastic) emission from small grains at 24$\mu$m. 
However,data from additional galaxies will be needed to 
confirm this. 

Finally, in the infrared SEDs predicted by Piovan et al. (2003) 
the $\langle K - 24 \rangle$ colors are expected to be 
enhanced in younger galaxies or in galaxies with a subcomponent  
of younger stars. 
When $\langle K - 24 \rangle$ for our sample galaxies are plotted 
against ages estimated from stellar indices, there is some 
hint of a negative correlation, but our sample is too 
small and the ages too uncertain to be confident. 

\section{Conclusions}

Our study of the mid-IR photometric properties of 
18 elliptical galaxies 
from 1.2$\mu$m to 24$\mu$m resulted in the following conclusions:

\noindent
(1) The surface brightness distributions at all photometric 
passbands between 1.2 and 24$\mu$m show good agreement with 
de Vaucouleurs $R^{-1/4}$ profiles. 
However, small deviations do exist among the passbands, 
resulting in significant radial color gradients.

\noindent
(2) The sample-averaged 
spectral energy distribution (SED) for elliptical 
galaxies peaks at $\sim1.6\mu$m 
near the $H$ band and exhibits a small variation with 
galactic radius.  
Emission at 5.8 and 8$\mu$m is systematically slightly lower 
near the galactic cores but emission at 24$\mu$m becomes stronger 
there.

\noindent
(3) The effective radius $R_e$ varies among 
various infrared passbands and has a sharp minimum at 2.15$\mu$m
in the $K$ band. 
$R_e$ at 24$\mu$m is also small but there is more variation 
among individual galaxies. 

\noindent
(4) Except for colors involving the 24$\mu$m passband, 
all radial sample-averaged color profiles 
$\langle i - j \rangle$ where $\lambda_i < \lambda_j$ 
have positive slopes within about $2R_e(K)$, 
suggesting a sensitivity to stellar metallicity 
which decreases with galactic radius. 
However, the slope variations of individual color 
profiles are somewhat enigmatic in detail. 
All color profiles involving the $K$ band, $\langle K - i \rangle$, 
have pronounced positive slopes.
For example the radial color profiles 
$\langle 3.6 - 4.5 \rangle$ and $\langle 4.5 - 8 \rangle$
are rather flat while 
$\langle K - 3.6 \rangle$ and $\langle K - 4.5 \rangle$ have 
significantly larger positive slopes.
The variation of $\langle K - 5.8 \rangle$ 
is comparable to that in $B - I$.
These results may suggest an 
anomalously enhanced emission in the $K$ band from metal rich stars,
possibly detracting from the common use of $K$ band
emission as an abundance-independent measure of local stellar mass.
Centrally enhanced $K$ band emission 
is also consistent with the minimum effective 
radius in this passband. 

\noindent
(5) The sample-averaged radial  
color variations for the various mid-IR passbands $i$ are consistent with 
the ratios of the half-light radii.  
For example, $\langle R_e(3.6)/R_e(4.5)\rangle$
is nearly unity and the mean radial color profile 
$\langle 3.6 - 4.5 \rangle$ is flat with radius. 
By contrast, $\langle R_e(5.8)/R_e(K)\rangle$ is large and 
the radial variation of $\langle K - 5.8 \rangle$ has 
a correspondingly larger positive slope. 

\noindent
(6) Colder grains that emit at longer wavelengths are generally 
more distant from their parent red giant star. 
Eventually, as grains move very far from the parent star, 
they are heated more by 
diffuse interstellar starlight or collisional 
impacts than by light from the parent star. 
Although the integrated
emission in the 24$\mu$m passband is mostly circumstellar 
(Temi, Brighenti \& Mathews 2007), 
we find that radial color profiles 
formed with the 24$\mu$m passband have uniquely peculiar shapes.
For example, our sample-averaged
$\langle K - 24 \rangle$ color profile has a maximum near
$0.5 R_e(K)$. 
This may indicate a transition to
interstellar heating for the grains emitting
at this long wavelength, but
our sample is too small to verify this with confidence.

\acknowledgements
This work is based on observations made with the Spitzer Space
Telescope, which is operated by the Jet Propulsion Laboratory,
California Institute of Technology, under NASA contract 1407.
Support for this work was provided by NASA through Spitzer
Guest Observer grant RSA 1276023.
Studies of the evolution of hot gas in elliptical galaxies
at UC Santa Cruz are supported by a Spitzer Theory Grant, 
NASA grants NAG 5-8409 \& ATP02-0122-0079 and NSF grant
AST-0098351 for which we are very grateful.

\clearpage

\begin{deluxetable}{rllrrc}
\tablecaption{Basic Properties of the Sample}
\tablewidth{10.cm}
\tablehead{
\colhead{NGC} &
\colhead{Type} &
\colhead{T\tablenotemark{a}} &
\colhead{$d$\tablenotemark{b}} &
\colhead{Log $L_B$\tablenotemark{b}} &
\colhead{2MASS?\tablenotemark{c}} \\
                   &
                   &
                   &
\colhead{(Mpc)} &
\colhead{($L_{B,\odot}$)} &
}
\startdata
 584 &  E4 & -4.6 &  23.76 & 10.42 & yes \\
2300 & Sa0 & -3.4 &  29.65 & 10.47 & yes \\
3377 &E4-5 & -4.8 &  10.71 & 9.78  & yes \\
3379 &  E1 & -4.8 &  10.71 & 10.12 & yes \\
3923 &E4-5 & -4.6 &  19.10 & 10.58 & yes \\
4365 &  E3 & -4.8 &  17.06 & 10.40 & yes \\
4406 &  E3 & -4.7 &  17.06 & 10.72 & yes \\
4472 &E2/S0& -4.8 &  17.06 & 10.96 & yes \\
4636 &E/S0 & -4.8 &  17.06 & 10.57 & yes \\
4696 &  E+1& -3.7 &  39.65 & 10.99 & yes \\
4697 &  E6 & -4.8 &  16.22 & 10.61 & yes \\
5044 &  E0 & -4.7 &  32.36 & 10.76 & no  \\
5322 & E3-4& -4.8 &  29.79 & 10.73 & yes \\
5557 &  E1 & -4.8 &  49.02 & 10.90 & no  \\
5813 & E1-2& -4.8 &  40.67 & 10.70 & yes \\
5831 &  E3 & -4.8 &  24.55 & 10.09 & no  \\
*5845 &  E  & -4.8 &  24.55 & 9.62  & no  \\
5846 & E0-1& -4.7 &  24.55 & 10.72 & no  \\
6703 & SA0 & -2.8 &  32.06 & 10.43 & no  \\
\enddata
\tablenotetext{a}{The morphological type T is taken from the HyperLeda
database.}
\tablenotetext{b}{Distances and Luminosities 
are calculated with $H_0=70$ km s$^{-1}$
Mpc$^{-1}$.}
\tablenotetext{c}{Availability of photometric data 
at the 2MASS website.}
\end{deluxetable}

\clearpage

\begin{deluxetable}{rccccccc}
\tablecaption{Mid-IR Photometric Data}
\tablewidth{10.cm}
\tablehead{ 
\colhead{} &
\colhead{} &
\colhead{} &
\colhead{} &
\colhead{} &
\colhead{} &
\colhead{} \\
}
\startdata
\cutinhead{Effective Radius $R_e(i)~({\prime}{\prime})$}
NGC & $K$ & $3.6\mu$m & $4.5\mu$m & $5.8\mu$m & $8\mu$m & $24\mu$m \\
    &     &           &           &           &         &          \\
\hline
584   & 23.64 & 25.71 & 27.10 & 29.15 & 28.80 & 23.82 \\
2300  & 22.78 & 22.02 & 21.62 & 23.93 & 22.66 & 25.34 \\
3377  & 25.48 & 29.50 & 29.58 & 31.88 & 30.65 & 27.57 \\
3379  & 29.82 & 35.06 & 34.06 & 34.56 & 36.30 & 22.87 \\
3923  & 40.25 & 47.67 & 47.76 & 48.66 & 49.78 & 47.01 \\
4365  & 37.78 & 44.22 & 44.20 & 47.38 & 47.85 & 41.67 \\
4406  & 59.13 & 62.10 & 62.30 & 58.53 & 58.28 & 52.32 \\
4472  & 55.86 & 64.73 & 63.92 & 59.48 & 67.88 & 51.79 \\
4636  & 56.29 & 62.61 & 61.73 & 62.92 & 61.65 & 58.90 \\
4696  & 40.51 & 46.53 & 47.88 & 48.46 & 50.05 & 49.99 \\
4697  & 39.58 & 45.16 & 50.02 & 56.49 & 53.59 & 29.72 \\
5044  &\nodata& 49.88 & 48.78 & 48.87 & 50.16 & 43.17 \\
5322  & 31.31 & 34.56 & 34.49 & 38.10 & 37.44 & 16.91 \\
5557  &\nodata& 24.07 & 25.58 & 25.44 & 24.07 & 15.15 \\
5813  & 36.17 & 43.31 & 42.6  & 52.60 & 47.10 & 22.71 \\
5831  & 19.43 & 20.67 & 21.03 & 23.77 & 23.76 &\nodata\\
%5845  & 5.12  &  5.01 &  5.09 &  6.40 &  5.95 &\nodata\\
5846  & 35.75 & 42.82 & 44.51 & 42.75 & 43.83 & 38.44 \\
6703  & 20.11 & 20.16 & 20.28 & 25.65 & 19.74 & 20.64 \\
\cutinhead{Surface Brightness $\mu_e(i)$ at $R_e(i)$ 
(mag $({\prime}{\prime})^{-2}$)}
NGC & $K$ & $3.6\mu$m & $4.5\mu$m & $5.8\mu$m & $8\mu$m & $24\mu$m \\
    &     &           &           &           &         &          \\
\hline
584   & 17.11 & 17.08 & 17.24 & 17.31 & 17.18 & 15.71 \\
2300  & 17.54 & 17.34 & 17.38 & 17.52 & 17.37 & 16.26 \\
3377  & 17.31 & 17.25 & 17.26 & 17.31 & 17.28 & 16.02 \\
3379  & 16.93 & 16.99 & 17.00 & 16.95 & 16.98 & 14.93 \\
3923  & 17.57 & 17.45 & 17.50 & 17.42 & 17.46 & 16.02 \\
4365  & 17.64 & 17.69 & 17.76 & 17.73 & 17.71 & 16.53 \\
4406  & 17.89 & 17.83 & 17.90 & 17.68 & 17.63 & 16.45 \\
4472  & 17.26 & 17.29 & 17.35 & 17.15 & 17.32 & 15.57 \\
4636  & 18.18 & 18.23 & 18.24 & 18.20 & 18.24 & 18.50 \\
4696  & 18.03 & 18.03 & 18.15 & 18.08 & 18.12 & 16.60 \\
4697  & 17.34 & 17.06 & 17.29 & 17.47 & 17.26 & 15.21 \\
5044  &\nodata& 18.70 & 18.74 & 18.25 & 18.23 & 16.98 \\
5322  & 17.46 & 17.57 & 17.60 & 17.67 & 17.60 & 14.79 \\
5557  &\nodata& 17.73 & 17.90 & 17.52 & 17.40 & 15.58 \\
5813  & 18.04 & 18.18 & 18.19 & 18.40 & 18.21 & 16.02 \\
5831  & 17.82 & 17.88 & 17.95 & 18.14 & 18.04 &\nodata\\
%5845  & 15.51 & 15.23 & 15.34 & 15.92 & 15.75 &\nodata\\
5846  & 17.60 & 17.94 & 18.06 & 17.87 & 17.89 & 16.35 \\
6703  & 17.85 & 17.72 & 17.76 & 18.04 & 17.58 & 16.09 \\
\cutinhead{Bandpass Luminosities $L(i)$ (erg s$^{-1}$)}
NGC & $K$ & $3.6\mu$m & $4.5\mu$m & $5.8\mu$m & $8\mu$m & $24\mu$m \\
    &     &           &           &           &         &          \\
\hline
584   & 43.01 & 42.61 & 42.32 & 42.09 & 41.90 & 40.89 \\
2300  & 43.07 & 42.69 & 42.40 & 42.15 & 41.94 & 40.88 \\
3377  & 42.26 & 41.92 & 41.66 & 41.40 & 41.19 & 40.04 \\
3379  & 42.73 & 42.37 & 42.09 & 41.83 & 41.63 & 40.40 \\
3923  & 43.14 & 42.79 & 42.51 & 42.30 & 42.07 & 40.87 \\
4365  & 42.99 & 42.62 & 42.34 & 42.11 & 41.92 & 40.60 \\
4406  & 43.20 & 42.82 & 42.54 & 42.27 & 42.07 & 40.86 \\
4472  & 43.48 & 43.12 & 42.83 & 42.55 & 42.40 & 41.13 \\
4636  & 43.07 & 42.70 & 42.41 & 42.17 & 41.96 & 40.67 \\
4696  & 43.52 & 43.18 & 42.91 & 42.66 & 42.48 & 41.26 \\
4697  & 43.05 & 42.70 & 42.45 & 42.23 & 42.05 & 40.68 \\
5044  &\nodata& 42.97 & 42.67 & 42.54 & 42.38 & 41.02 \\
5322  & 43.26 & 42.90 & 42.63 & 42.40 & 42.22 & 41.10 \\
5557  &\nodata& 43.08 & 42.80 & 42.64 & 42.43 & 41.08 \\
5813  & 43.19 & 42.86 & 42.57 & 42.39 & 42.14 & 40.77 \\
5831  &\nodata& 42.33 & 42.05 & 41.80 & 41.61 &\nodata \\
%5845  &\nodata& 41.93 & 41.64 & 41.41 & 41.24 &\nodata \\
5846  & 43.18 & 42.88 & 42.61 & 42.36 & 42.16 & 40.99 \\
6703  &\nodata& 42.63 & 42.36 & 42.15 & 41.88 & 40.88 \\
\enddata

\end{deluxetable}

%\end{document}

\clearpage
%\vskip1.in
\begin{figure}[ht]%1 
%\figurenum{1}
\centering
%\vskip2.in
%%\includegraphics[bb=90 216 522 569,scale=0.9,angle= 270]
%\includegraphics[bb=90 166 522 519,scale=1.0,angle= 0]
\includegraphics[bb=50 216 422 669,scale=0.8,angle=0]{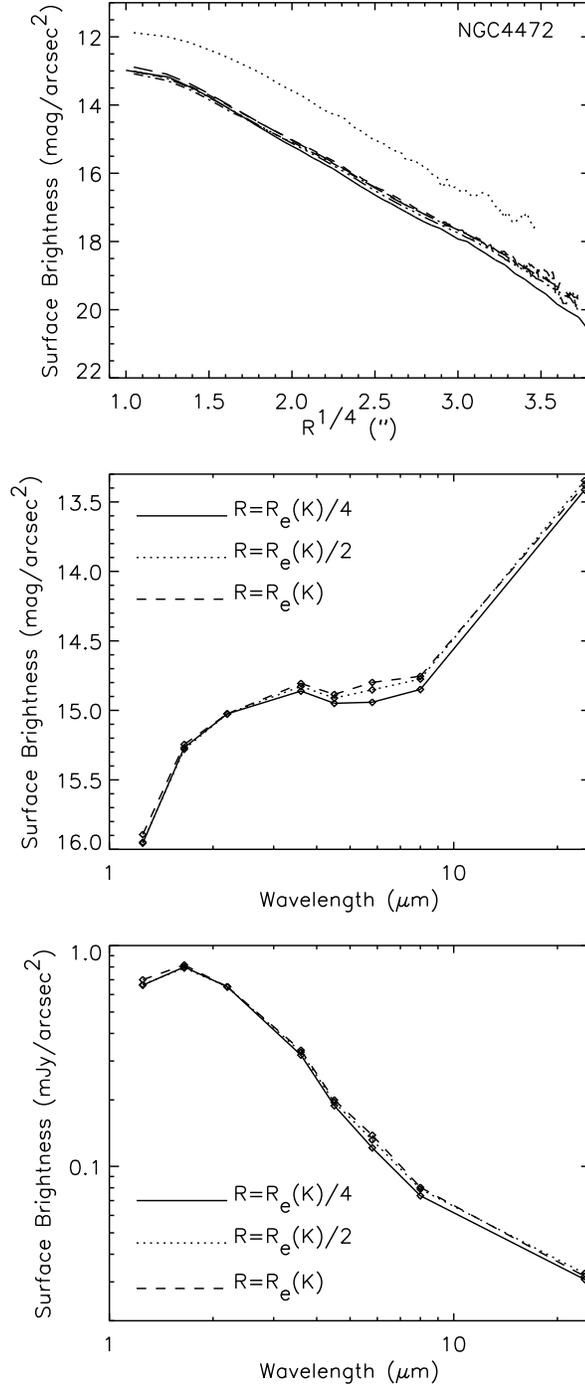}
\vskip2.3in
\caption{
Mid-infrared surface brightness and spectral energy distributions 
(SEDs) for NGC 4472.
{\it Top panel:} surface brightness profiles at six mid-infrared
passbands: $K$ band (solid line), IRAC 3.6, 4.5, 5.8 and 8.0$\mu$m 
bands (having different line types but strongly overlapping), 
and MIPS 24$\mu$m (dotted line).
{\it Central panel:}
Spectral energy distributions $\mu_i$ 
and evaluated at three different radii in units 
of magnitude arcsec$^{-2}$. The data points refer from left to 
right to the $i = J$, $H$, $K$, 3.6, 4.5, 5.8, 8.0 and 24$\mu$m passbands.
The SEDs are arbitrarily normalized to the $K$ band.
{\it Bottom panel:}
Identical to the central panel except the surface brightness 
is given in mJy arcsecond$^{-2}$.
}
\label{f1}
\end{figure}

\clearpage
\begin{figure}[ht]%1
%\figurenum{2}
\centering
\vskip1.in
\includegraphics[bb=50 166 422 619,scale=0.8,angle=0]{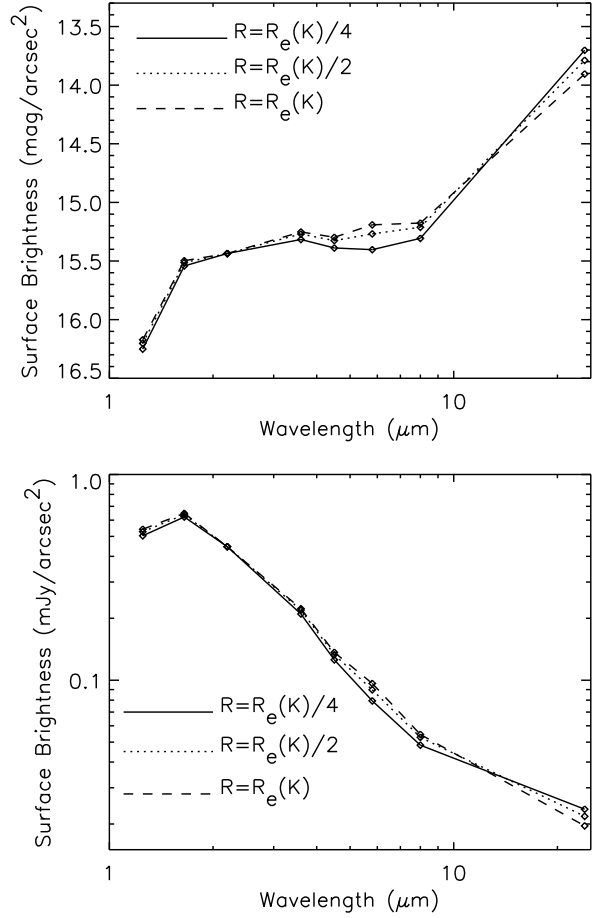}
\vskip2.in
\caption{
Sample-averaged SEDs at three radii normalized with $R_e(K)$ 
and normalized to the $K$ band as in Figure 1.
{\it Top panel:} mean SEDs in units of magnitude arcsec$^{-2}$.
{\it Bottom panel:} mean SEDs in units of mJy arcsec$^{-2}$.
}
\label{f2}
\end{figure}

\clearpage
\vskip2.in
\begin{figure}[h]%1
%\figurenum{3}
\centering
\vskip3.in
\includegraphics[bb=100 166 472 619,scale=0.7,angle=90]{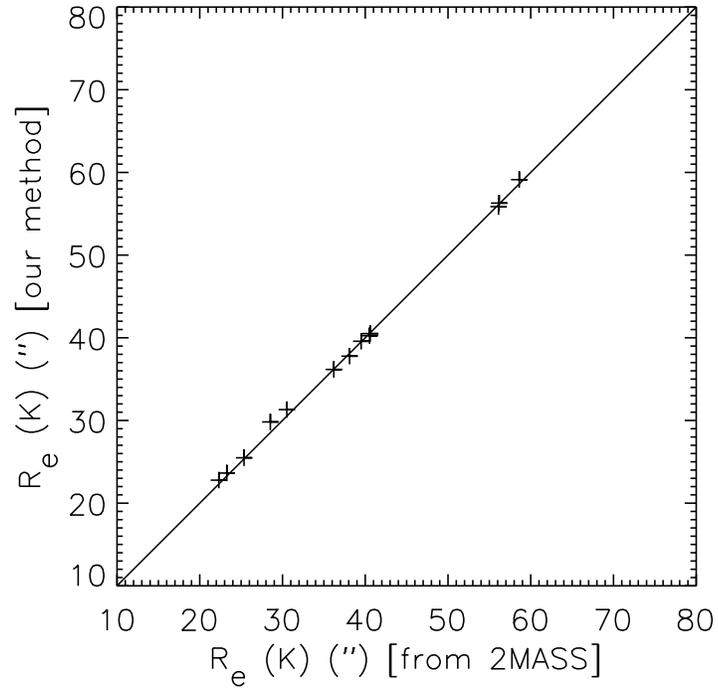}
\vskip0.7in
\caption{
Comparison of the $K$ band effective radius of sample galaxies 
measured directly from 2MASS images using our photometric procedure 
with $R_e(K)$ found by Jarrett (2003) using the 2MASS procedure. 
}
\label{f3}
\end{figure}

\clearpage
%vskip1.in
\begin{figure}[ht]%1
%\figurenum{4}
\centering
\vskip2.in
\includegraphics[scale=0.8,angle=90]{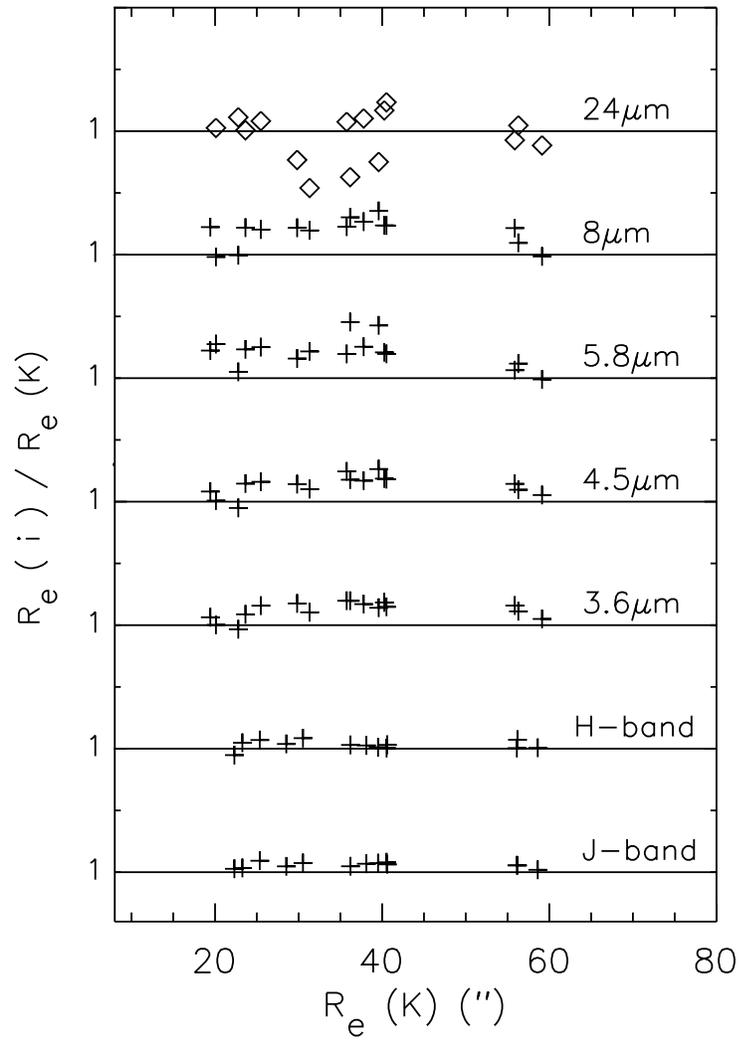}
%\vskip.7in
\caption{
Ratios of the effective radius at seven passbands $R_e(i)$ 
to the $K$ band effective radius. 
$R_e(24)/R_e(K)$ ratios are shown with open diamonds. 
}
\label{f4}
\end{figure}

\clearpage
%\vskip1.in
\begin{figure}[ht]%1
%\figurenum{5}
\centering
%\vskip3.in
%%\includegraphics[bb=90 216 522 569,scale=0.9,angle= 270]
%\includegraphics[bb=90 166 522 519,scale=1.0,angle= 0]
\includegraphics[bb=100 166 472 619,scale=0.8,angle=90]{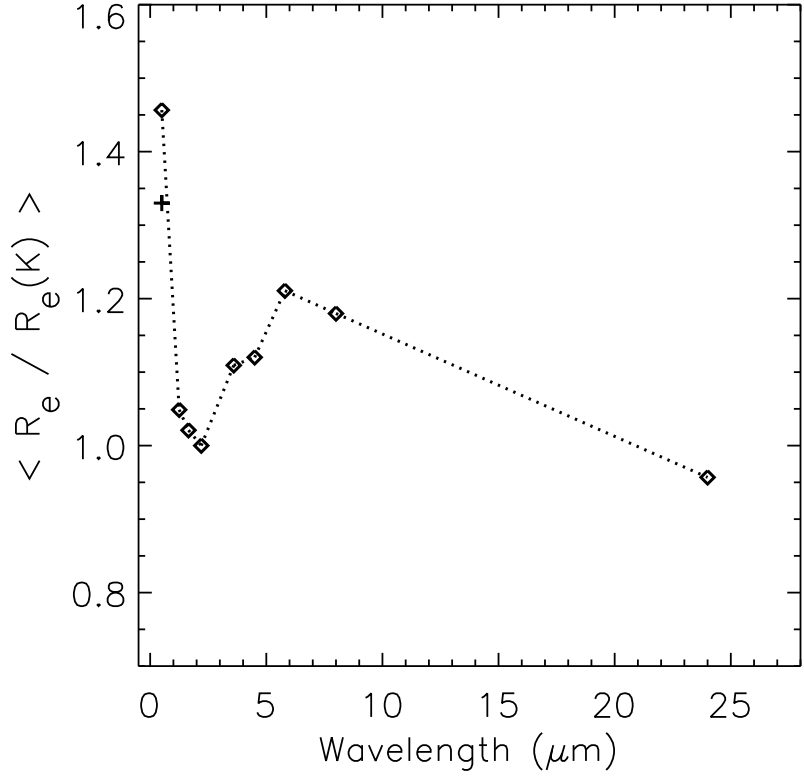}
\vskip.7in
\caption{
Variation with mean passband wavelength 
of sample-averaged ratios of the effective radii 
in each passband to that in the $K$ band. 
Minima occur at the $K$ band (2.15$\mu$m) and at 24$\mu$m. 
The + symbol shows the ratio 
$\langle R_e(V)/R_e(K)\rangle$, 
for 273 early type galaxies (Ko \& Im 2005).
}
\label{f5}
\end{figure}

\clearpage
\vskip1.in
\begin{figure}[ht]%1
%\figurenum{6}
\centering
\vskip2.in
\includegraphics[bb=200 166 572 619,scale=0.7,angle=90]{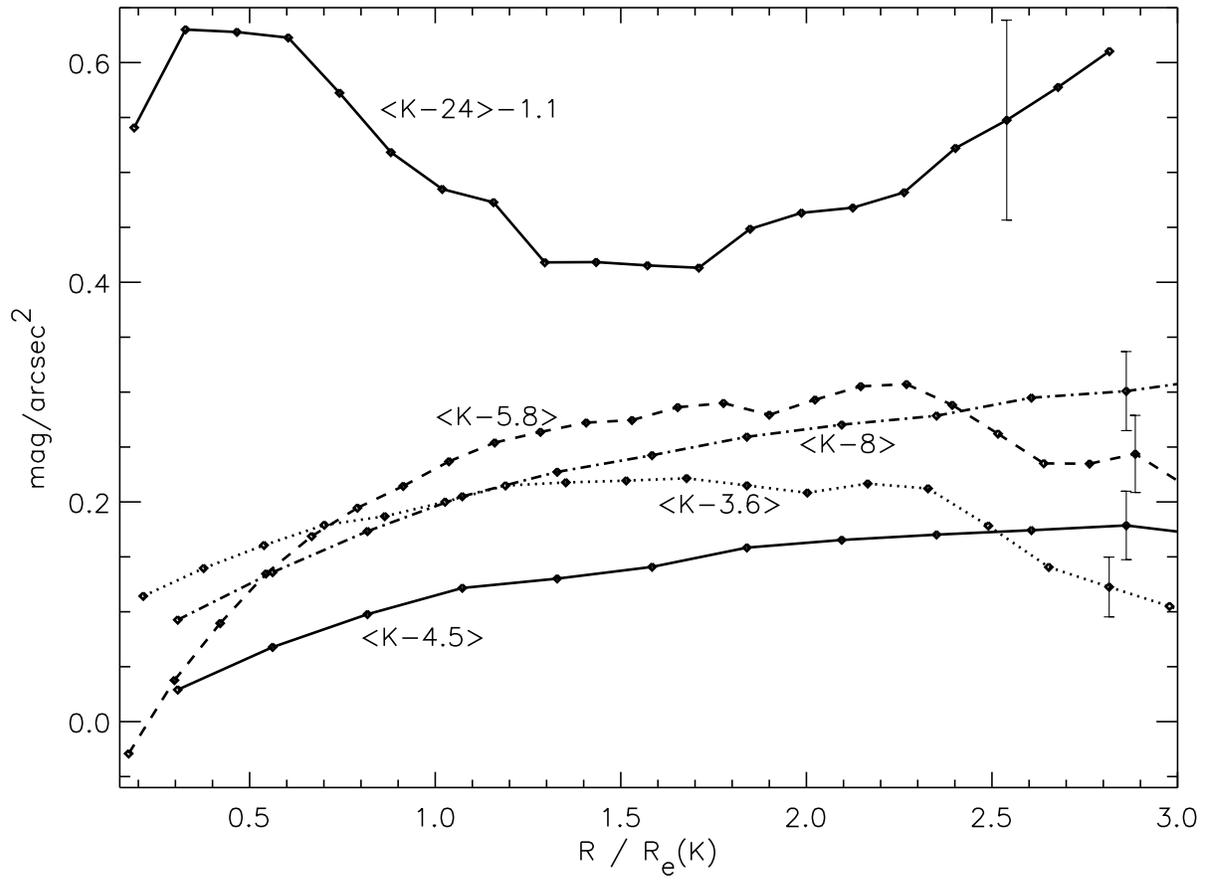}
\vskip1.5in
\caption{
Variation of five sample-averaged radial color profiles 
(in surface brightness units)
with galaxy radius normalized with $R_e(K)$.
}
\label{f6}
\end{figure}

\clearpage
\vskip4.in
\begin{figure}[ht]%1
%\figurenum{7}
\centering
\vskip2.in
\includegraphics[bb=200 166 572 619,scale=0.7,angle=90]{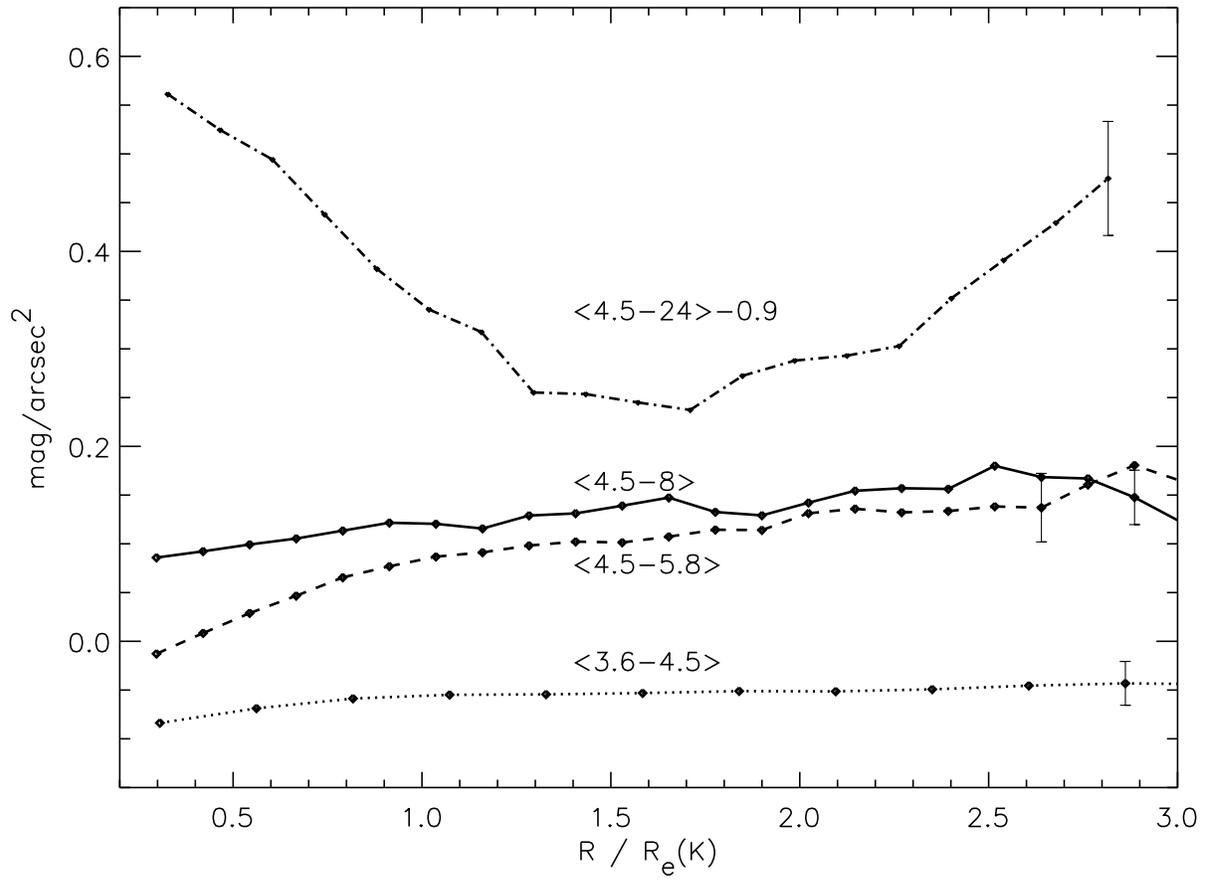}
\vskip1.5in
\caption{
Variation of four additional sample-averaged radial color profiles
(in surface brightness units)
with galaxy radius normalized with $R_e(K)$.
}
\label{f7}
\end{figure}

\clearpage
\vskip4.in
\begin{figure}[ht]%1
%\figurenum{8}
\centering
\vskip1.in
\includegraphics[bb=50 166 422 619,scale=0.9,angle=0]{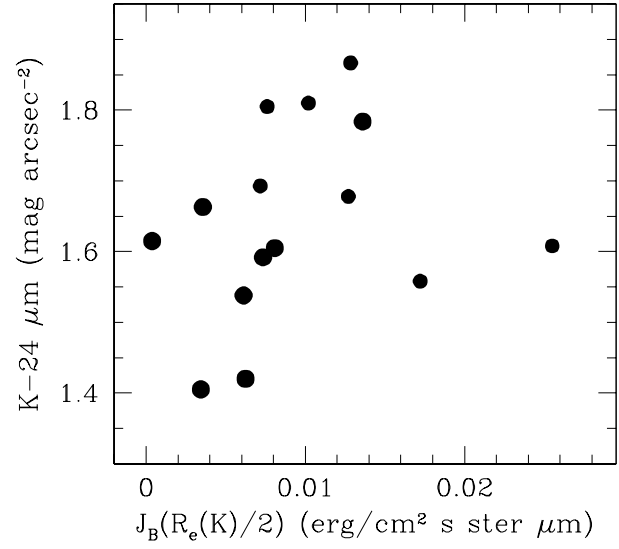}
%\vskip.7in
\caption{
Plot showing the ($K$ - 24$\mu$m) color evaluated at $R_e(K)/2$ 
with the mean intensity of $B$ band diffuse galactic light 
at the same radius.
The two outlying galaxies at the right are NGC 3379 
($J_B$ = 0.017)
and NGC 584 ($J_B$ = 0.0255). 
%Galaxies with flat morphologies (E4-E6 or S0) are noted with open 
%circles. 
}
\label{f8}
\end{figure}

\clearpage
\appendix
\section{Spatial Resolution of 2MASS and IRAC Passbands}

In our determination of half light radii and 
color profiles for elliptical galaxies
we have not deconvolved the surface brightness images 
to correct for the point spread function (PSF).
In this Appendix we justify this procedure by showing that
the widths of the PSFs are small compared to the K-band effective radius 
which characterizes the scale of 
the metallicity gradients of our sample galaxies. 

Figure 1A shows the azimuthally averaged PSF for the 2MASS 
$K_s$ and IRAC bands, all normalized to unity at the center.
The FWHM of all IRAC bands are less than 2$^{\prime\prime}$
and the FWHM of the $K_s$ band is about 2.5$^{\prime\prime}$, 
only slightly larger.
These PSFs are all much less than the half light radii of 
our sample galaxies listed in Table 2, 
$R_e(K) \approx 20 - 55^{\prime\prime}$. 
The PSFs alter the infrared surface brightness distributions 
only in the central few arcseconds 
which we do not discuss in this paper. 
Our implicit assumption that the redistribution of scattered light
does not vary across
the IRAC fields of view is supported by a recent study of the
surface brightness of the elliptical galaxy NGC 5044 when the galactic
center is placed in various non-central locations in the 
field of view.\footnote{
http://spider.ipac.caltech.edu/staff/jarrett/irac/calibration/droop/}

The IRAC website\footnote{
http://ssc.spitzer.caltech.edu/irac/calib/extcal/}
at the {\it Spitzer Science Center} describes 
in some detail how to determine the surface brightness of 
extended objects such as elliptical galaxies 
and the various difficulties that can arise.
An upper limit
on the effect of scattered light on surface brightness
color variations can be estimated, as suggested at this website, 
by using a double PSF convolution.
Our (properly calibrated) surface brightness data $\mu(i)$
for each passband is convolved with the PSF$(i)$ for that passband. 
To evaluate the (maximum) influence of scattered light 
on the color $\mu(i) - \mu(j)$, we convolve $\mu(i)$ with PSF$(j)$
and $\mu(j)$ with PSF$(i)$, i.e. a double convolution.

The solid lines in Figure 2A show the binned surface brightness 
color profiles ($K - 5.8\mu$m) and (4.5 - 5.8$\mu$m) 
for elliptical galaxy NGC 6703 
in which each passband is slightly broadened by the PSF of that band. 
These profiles are similar to the surface brightness color 
profiles discussed in Figures 6 and 7,
except in this example they refer to a single 
typical galaxy, NGC 6703, not the sample average. 
The dashed lines in Figure 2A show the doubly convoluted color profiles 
in which the image in each passband is degraded by the PSF of both bands. 
It is clear that the radial color variations are essentially unchanged 
by the double PSF convolution 
except at $R/R_e(K) \lta 0.2$ which we do not consider 
in our discussion. 
We conclude that our half light radii and 
radial color profiles are essentially unaffected by the 
2MASS or IRAC PSFs.

At the MIPS 24$\mu$m wavelength the PSF is somewhat larger, 
about 5'' FWHM. 
Nevertheless, we assume that this broadening 
also has little influence on 
the half light radius or color profiles involving the 24$\mu$m 
passband, both of which vary on larger scales comparable to $R_e$.

\clearpage
\begin{figure}%1
\centering
\vskip2.in
\includegraphics[bb=90 166 522 519,scale=0.8,angle= 90]
{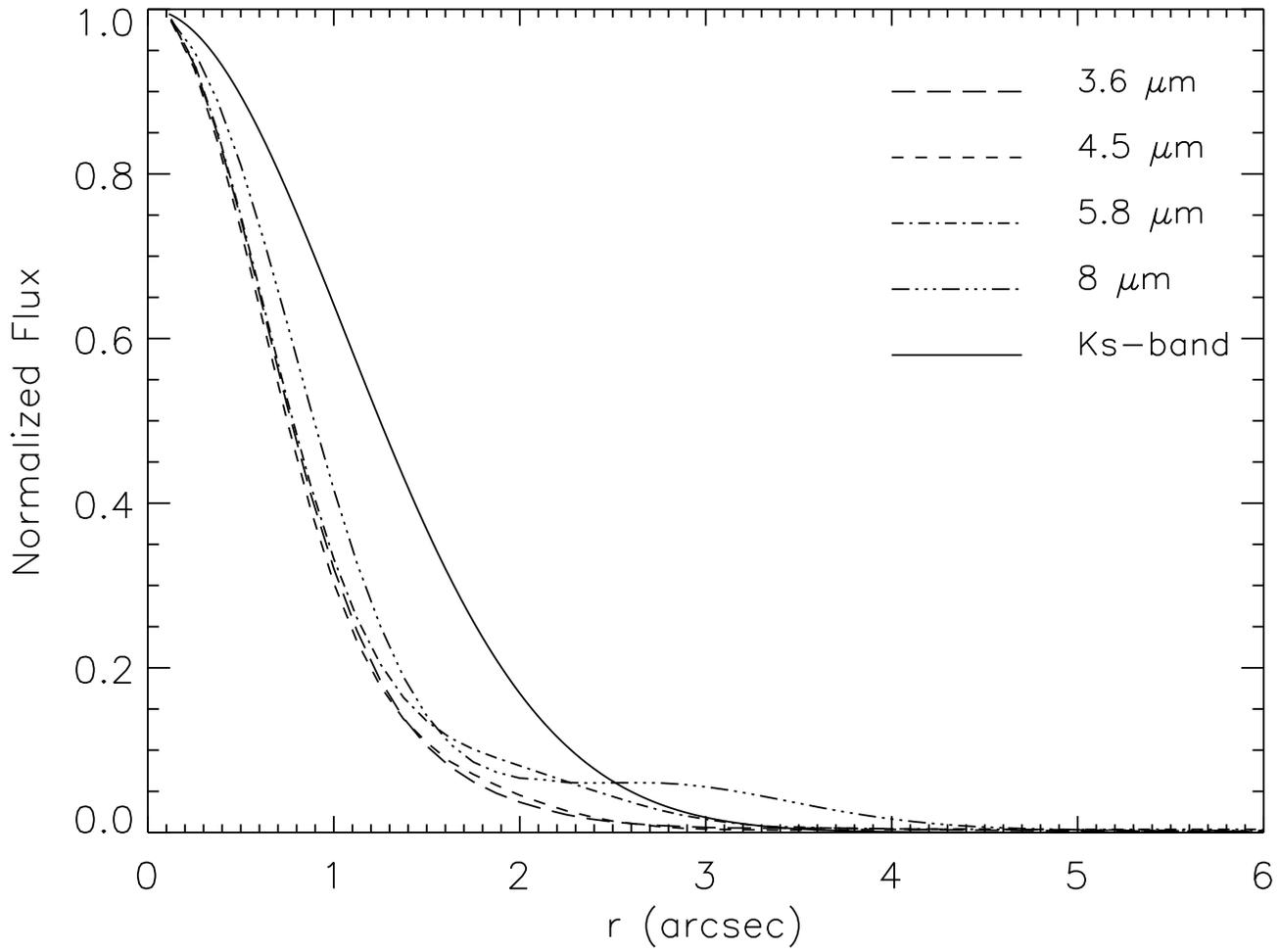}
\vskip.7in
\caption{
Azimuthally averaged
point spread functions for the $K_s$-band and the four IRAC bands.
}
\label{f1A}
\end{figure}

\clearpage
\begin{figure}%2
\centering
\vskip2.in
\includegraphics[bb=90 166 522 519,scale=0.8,angle= 90]
{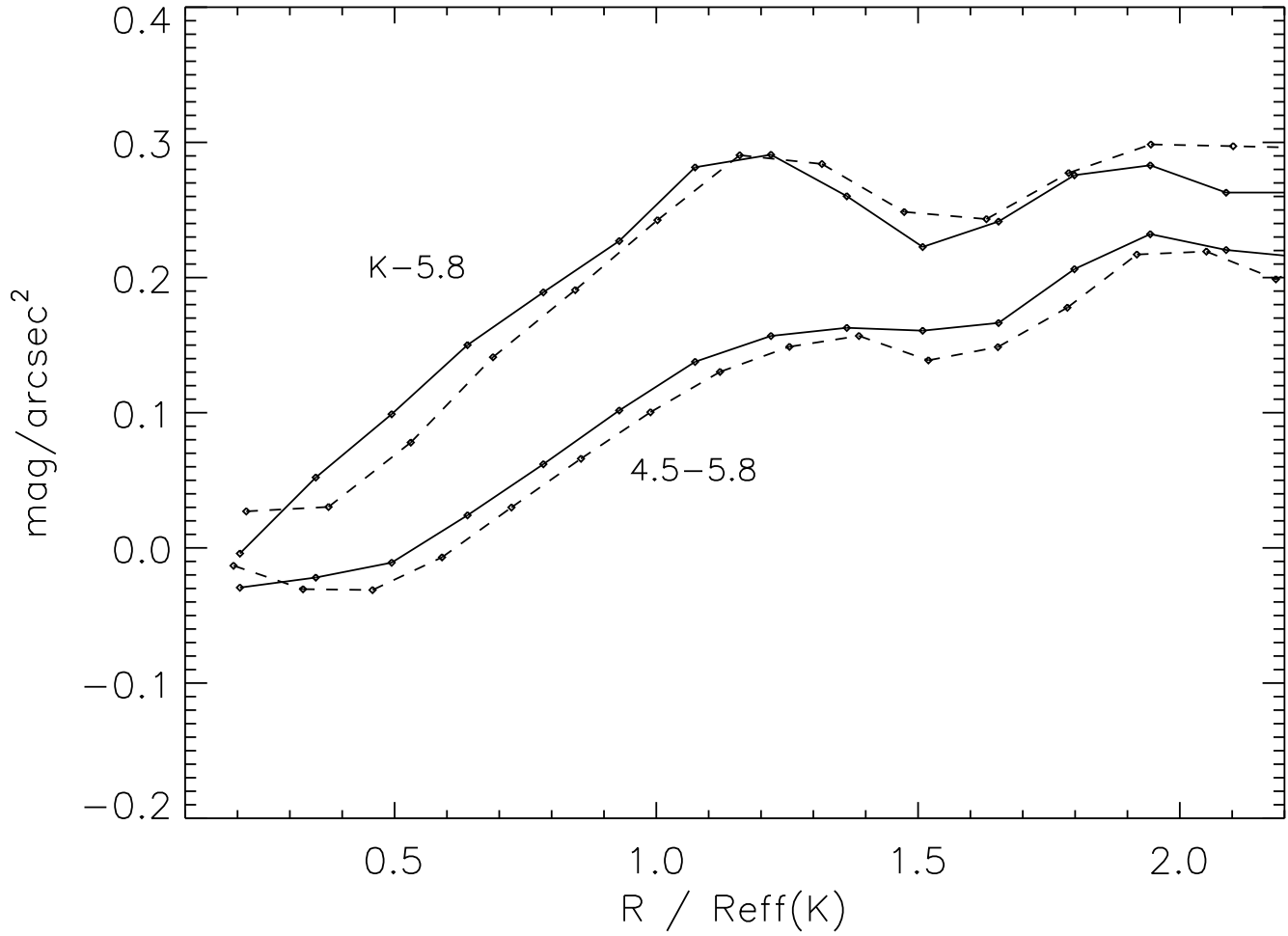}
\vskip.7in
\caption{
Color profiles for the galaxy NGC 6703 for singly convolved
(solid lines) and doubly convolved (dashed lines) point spread
functions.
}
\label{f2A}
\end{figure}

\end{document}